\documentclass{elsart}

\usepackage{natbib}  
 
\usepackage{graphicx}

\begin{document}


\runauthor{Ros, Zensus and Lobanov}


\begin{frontmatter} 
\title{Multiband polarimetric and total intensity imaging of 3C\,345}
\author[MPIfR]{E. Ros}
\author[MPIfR]{J.A. Zensus}
\author[MPIfR]{A.P. Lobanov}  



\address[MPIfR]{Max-Planck-Institut f\"ur Radioastronomie, 
Bonn, Germany}
 

\begin{abstract} 
We monitored the superluminal QSO 3C\,345 at three epochs during a one-year
period in 1995--1996,
observing with the VLBA at 22, 15, 8.4, and 5\,GHz.  We imaged the radio 
source both in total and in polarized intensity.  
In the images at 5 and 8.4\,GHz, the jet emission is traced up to 20 
milliarcseconds (mas) from the jet core.  In the 15 and 22\,GHz
images, we identify several enhanced emission regions moving at apparent speeds
of 5$c$.
Images of the linear polarized emission
show predominantly an alignment of 
the electric vector with the extremely curved jet along the inner 
part of the high frequency
jet.  At 5\,GHz, the jet shows remarkably strong fractional polarization 
($m\sim15\%$) with the electric vector perpendicular to the jet orientation.
\end{abstract} 



\end{frontmatter} 

\section{Introduction}
\label{sec:intro} 

The QSO 3C\,345 ($V$=16$^{\rm mag}$, $z$=0.595) 
is one of the best examples of a radio source jet in a 
core-dominated radio source.  It has been monitored with VLBI since
1979 \citep{unw83,bir86,bro94,war94,ran95,zen95a,zen95b,lob96,tay98}. 
On arcsecond scales the 
source contains a compact region at the base of a 4$^{\prime\prime}$ jet that
is embedded in a diffuse steep-spectrum halo \citep{kol89}.  

>From astrometric measurements, the
parsec-scale core has been shown to be stationary within
uncertainties of 20\,$\mu$as\,yr$^{-1}$ \citep{bar86}.
The parsec-scale jet consists of several prominent enhanced emission regions
(jet components) apparently ejected at different position angles (P.A.\
ranging from 240$^\circ$ to 290$^\circ$)
with respect to the jet core.  The components move along curved
trajectories that can be modelled by a simple helical jet.
The curvature of the trajectories may be caused by 
some periodic process at the jet origin, like orbital
motion in a binary black hole system or Kelvin-Helmholtz 
instabilities \citep{ste95,qia96,har87}.

3C\,345 was observed with VLBI during the first half of the
90's almost every 6 months at different frequencies.  The
component motions were well monitored through these observations
(cf.\ \citet{lob96}) allowing precise determination of component motions.  
The present work extends this effort with an enhanced data set at four
frequencies, including polarization data.

Assuming a 
standard Friedmann cosmology, with $H_0$=100$h$\,km\,s$^{-1}$\,Mpc$^{-1}$
and $q_0$=0.5, an angular scale of 1\,mas corresponds to 3.79\,$h^{-1}$\,pc
for a cosmological redshift of 0.595.  An apparent angular motion of
1\,mas\,yr$^{-1}$ represents an apparent speed of 
$\beta_{\rm app}$=19.7\,$h^{-1}$.  

\section{Observations and Imaging.
\label{sec:obs_img} }

We observed 3C\,345 with the VLBA at 3 epochs (1995.84, 1996.41,
and 1996.81) at 22, 15, 8.4, and 5\,GHz, using a bandwidth of 16\,MHz.
At each frequency, the source was observed for about
14\,hrs, using 5-minute scans and interleaving all observing frequencies.
Some calibrator scans (on 3C\,279, 3C\,84, NRAO\,91, OQ\,208, and 3C\,286)
were inserted during the observations. The data
were correlated at the NRAO\footnote{The National
Radio Astronomy Observatory (NRAO) is operated by Associated
Universities, Inc., under cooperative agreement with the National
Science Foundation.} VLBA correlator in Socorro, New Mexico, USA.

\subsection{Total Intensity Images.\label{subs:totint}}

The data were fringe-fitted and calibrated in AIPS and 
imaged using the differential mapping program DIFMAP \citep{she94}.
High dynamic range images (1:1000) were obtained for
all epochs and frequencies.  At the higher frequencies (15 and
22\,GHz), we identified the core D and two main components 
(tentatively identified as C8 and C7) in
the inner 3\,mas \citep{ros99}.  The components travel outwards
with respect to the core, with observed
apparent superluminal speeds of about 5$c$.
At 5 and 8.4\,GHz the jet extends up to 20\,mas distances from the core.

\begin{figure}[hbtp] 
\includegraphics[scale=0.73,angle=0]{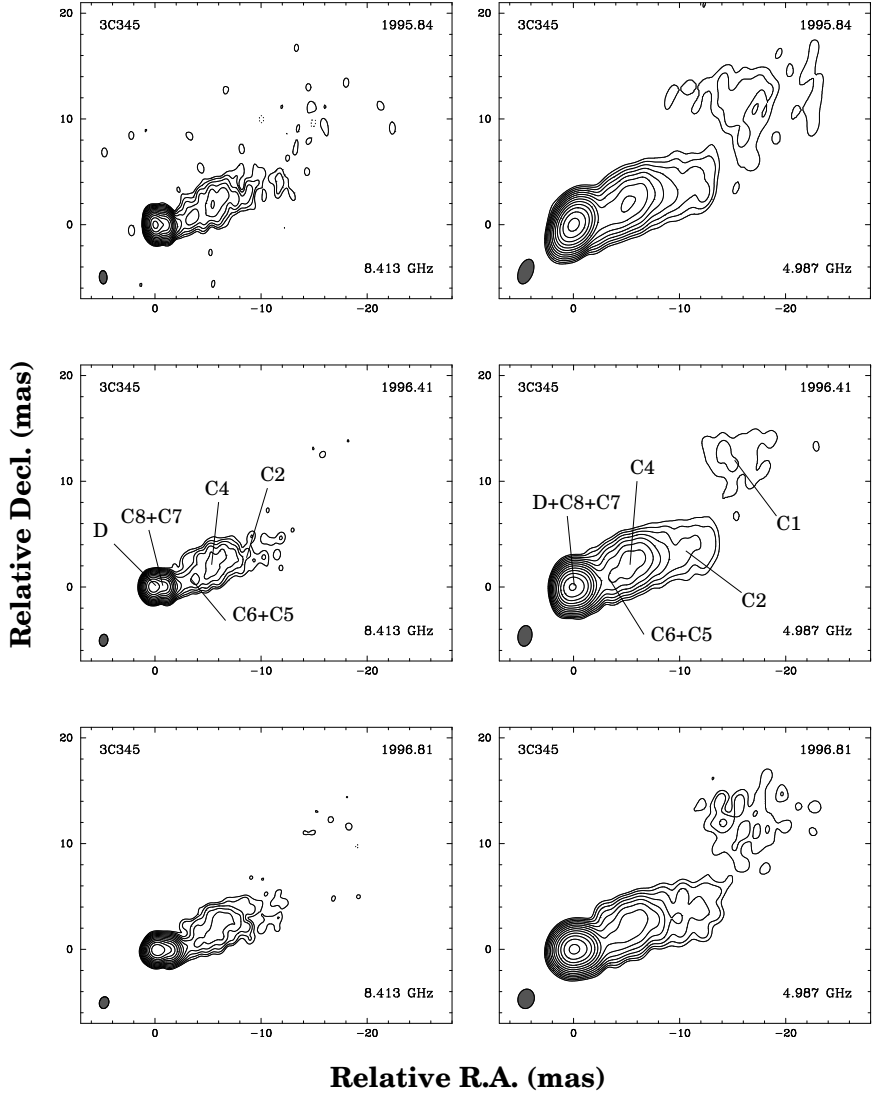}  
\caption{Total intensity VLBA images of 3C\,345 at 8.4 and 5\,GHz for 
epochs 1995.84, 1996.41, and 1996.81.  We can identify the
components D, C8, C7, C6, C5, C4, C2, and C1 in these images.
}
\label{fig:maps-8.4-5} 
\end{figure}  

Fig. \ref{fig:maps-8.4-5} shows the images at 8.4 and 5\,GHz for all 
3 observing epochs.
Model fitting of Gaussian profile components 
to the jet at 5 and 8.4\,GHz
includes the inner components seen at the higher frequencies, 
and additionally more extended components of the jet further out.  
The proper motions of these regions are typically of $\sim$4.5$c$. 
At $\sim$6.2\,mas from the core (P.A.\ $\sim -70^\circ$) the emission 
can be fitted for the six images with a component, having a total
flux density of $\sim$0.6\,Jy at 8.4\,GHz and $\sim$0.7\,Jy at 5\,GHz.
The proper motion for this component is $\sim$6$c$.
It is most plausibly identified with the C4 component discussed by, e.g.,
\citet{lob96}.
Component C2 is fitted at a position $\sim$12\,mas away from the core 
(P.A.\ $\sim -70^\circ$), with a flux density of $\sim$0.1\,Jy both
at 8.4 and 5\,GHz.  The extended nature of this component makes proper
motion determinations very uncertain.  At 5\,GHz another component, C1,
separated
$\sim$20\,mas from the core (P.A.\ $\sim -55^\circ$), with a 
flux density above 
0.2\,Jy for all three epochs, permits us to trace the northern extended jet 
emission (the FWHM of the elliptical component is about 10\,mas).
A comparison with previous model-fitting results will provide 
a better understanding of the kinematic properties of the
jet.

\subsection{Polarized Images.
\label{subs:polimg} }

Recently, important progress has been done in the VLBI polarimetric
observations.  The polarization calibration at high frequencies is
now possible using the target source itself as a calibrator \citep{lep95}.
This technique is non-iterative and is insensitive to structure in the 
polarization calibrator.  
We have applied this approach to determine the instrumental polarization
of the observing stations and to image the linear polarization in 3C\,345.
We do not apply Faraday rotation to the observed electric
vector position angles presented here, relying on the results of 
\citet{tay98}, who reports small rotation measurements in 3C\,345 at
frequencies higher than 5\,GHz.

\begin{figure}[hbt]
\centering
\includegraphics[scale=0.57,angle=0]{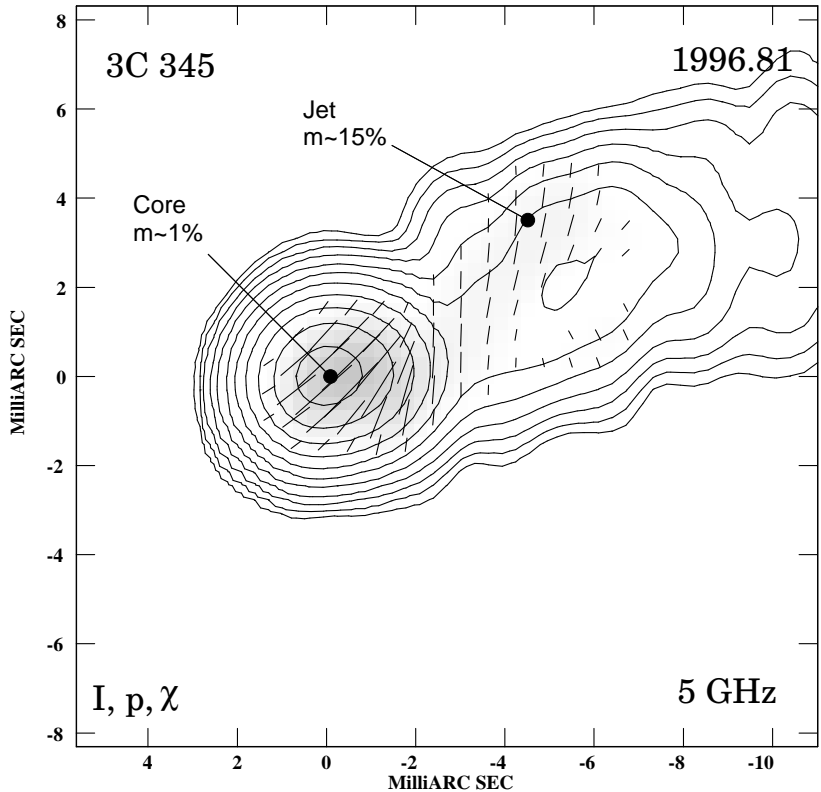}  
\caption{Polarized intensity electric vectors ($\chi$) overlaid on total
intensity ($I$) contours and grey scale polarized intensity ($p$) images
for 3C\,345. ($P=Q+iU=pe^{2 i \chi}=mIe^{2 i \chi}$, where $Q$ and $U$
are  Stokes parameters, $p=mI$ is the polarized linear intensity,
$m$ is the fractional linear polarization, and $\chi$ is the
position angle of the electric vector in the sky).  It is obvious that
the electric vector is almost perpendicular to the jet at core separations
from 3 to 7\,mas.}
\label{fig:c-pol}
\end{figure}

In \citet{ros99} we show some polarization results of this monitoring
at 22\,GHz, where
the electric vector is roughly parallel to the jet direction in the inner
part of the jet.
\citet{lis98} report similar vector alignment with the jet at 43\,GHz, for
some blazars in the inner jet regions.
At lower frequencies, the distribution of the polarized emission differs 
significantly from that seen at 22 and 15\,GHz.
The core is less polarized ($m$$\sim$1\%) than at higher frequencies,
showing an electric vector oblique to the jet direction.  Both
facts may be caused by the blending of differently oriented polarized
components.
The degree of polarization at 5\,GHz in the jet reaches values
over 15\%.  We present an image of the third epoch of our monitoring
in Fig.\ \ref{fig:c-pol}.  Both previous epochs display very similar
features for the jet. 

These findings seem to be consistent with the conclusions of
\citet{bro94}, who reported, at 5\,GHz, a weakly polarized 
core in 3C\,345 and a fractional polarization reaching 15\% in the jet.  
The electric vector position angles were 
found to be variable in the core from one epoch to the other but 
perpendicular to the jet for all epochs. 
The images obtained by \citet{tay98} at 8.4\,GHz are similar to
our results, and also imply a magnetic field aligned with the
jet direction at 3-10\,mas distances from the core.
\citet{caw93} suggested that 
the longitudinal component of the magnetic 
field can increase with the distance from the core as a result of
shear from the dense emission line gas near the nucleus.  At larger
distances from the core 
the shocks may be too weak to dominate the underlying field, resulting in
the observed electric field perpendicular to the jet.


\section{Conclusions}

We have studied the QSO 3C\,345 with the VLBA at three epochs 
and four frequencies, 
analyzing the properties of its total intensity and polarized 
emission.  We have monitored the kinematics of the radio source, 
identifying components and proper motions. 
The polarized intensity images show  the alignment
between the jet direction and the
electric vector position angle in the inner regions of the jet at
high frequencies, and the orthogonality of the electric vector to
the jet direction in the outer regions of the jet.  These results
confirm reliably the previously reported 
observations \citep{bro94,lep95,tay98}
covering different frequencies and jet regions in 3C\,345.

Continuing the multiband monitoring of 3C\,345 with VLBI, enhanced
by regular polarimetric studies, can provide
important clues on the physics of the parsec-scale jet.
Higher frequencies
and better resolution observations (as those provided by orbital VLBI results
with HALCA) should help to better constrain the models of this QSO.

\end{document}